\documentclass[12pt,a4paper]{article}
\usepackage[utf8]{inputenc}
\usepackage[normalem]{ulem}

\title{A Maxwell Fish-Eye Lens in a Bose-Einstein Condensate}

\author{Jelte Duchêne$^{1,*}$, Elinor Kath$^1$, Floriane Arrouas$^1$, Hanyi Jang$^1$, \\Helmut Strobel$^1$, Markus K. Oberthaler$^1$, Jay Mehta$^2$, Liam M. Farrell$^2$, \\Wyatt Kirkby$^{1,3}$, and Duncan H.J. O'Dell$^2$}

\date{$^1$ Kirchhoff-Institut für Physik, Universität Heidelberg, Heidelberg, Germany\\
$^2$Department of Physics and Astronomy, McMaster University, 1280 Main St. W., Hamilton, ON, Canada L8S 4M1\\
$^3$ Physikalisches Institut, Universität Heidelberg, Heidelberg, Germany\\
[1ex]
E-mail: \href{mailto:maxwellfisheyelens@matterwave.de}{maxwellfisheyelens@matterwave.de}\\
[2ex]
    }

\usepackage[
backend=biber,
style=numeric,
sorting=none,
]{biblatex}

\usepackage{float}
\usepackage{hyperref}
\usepackage{graphicx}
\usepackage{amsmath}
\usepackage{amssymb}
\usepackage{hyperref}
\usepackage{subcaption}
\usepackage{xcolor}
\usepackage{dsfont}
\usepackage{braket}
\usepackage{physics}
\usepackage{soul}
\usepackage{bm}
\usepackage{mathrsfs}

\usepackage[margin=2.5cm]{geometry}

\numberwithin{equation}{section}
\numberwithin{footnote}{section}

\begin{document}

\maketitle
\begin{abstract}
\noindent
We experimentally realize an analogue of the optical Maxwell fish-eye lens (MFEL) using phononic excitations in a Bose–Einstein condensate (BEC). A MFEL is characterized by a radially symmetric, spatially varying refractive index with the remarkable property that rays emitted from any point within the lens are perfectly focused at their antipodal points. While the implementation of such gradient-index lenses is challenging in conventional optical systems, BECs offer a highly tunable platform in which the spatially varying speed of sound of collective excitations -- phonons, the acoustic analogues of photons -- can be engineered and their dynamics observed in real time. Time-resolved measurements of phonon wavefronts reveal focusing behavior that shows good agreement with analytical theory and numerical simulations. This work provides both a geometric and physical framework for engineering effective refractive indices using ultracold atoms, and simulating wave propagation on effective spherical geometries.
\end{abstract}

\section{Introduction}

\begin{figure}[!t]
    \centering
	\begin{minipage}{1\textwidth}
		\centering
		\includegraphics[width=\linewidth]{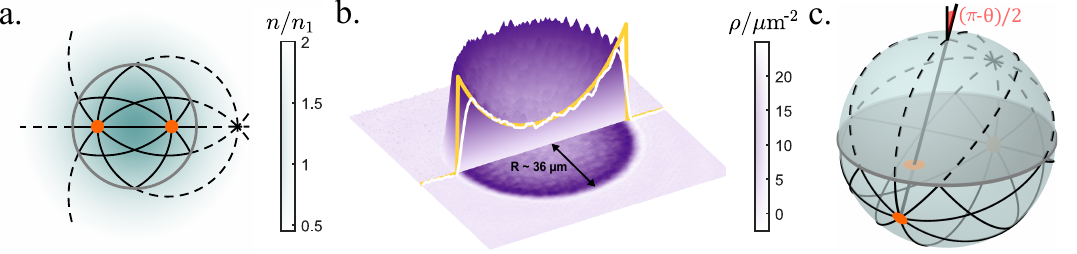}
	\end{minipage}\\[-0.2cm]  \caption{\textbf{a.}~Schematic diagram of the refractive index for a 2D Maxwell fish-eye lens, showing light rays emanating from a particular point and perfect focusing at the corresponding antipodal point. The solid and dashed black lines respectively represent the trajectories with and without the presence of a reflecting mirror at $r=R$ (solid gray line). \textbf{b.}~Corresponding experimental condensate atomic density profile. The white (experiment) and yellow (theory) lines are 1D slices, and separate a 3D from a 2D depiction of the density. \textbf{c.}~Light rays (great circles) on the surface of a sphere with constant refractive index $n_{1}$, whose stereographic projection onto the 2D equatorial plane corresponds to the ray trajectories in the 2D lens. The equator plays the role of the mirror in this representation.}
    \label{fig:MFEL_rays}
\end{figure}

The Maxwell fish-eye lens (MFEL) serves as an archetype for perfect, aberration-free optical imaging. It was introduced by Maxwell, who deduced that a radially dependent refractive index of the form
\begin{equation}
    n(r) = \frac{2n_1}{1+\left(r/R\right)^2}
    \label{eq:n_MFEL_pre}
\end{equation}
causes light rays to propagate along closed circular paths, leading to perfect point-to-point imaging in the framework of geometrical optics~\cite{Maxwell_1990}. Here, $r$ denotes the radial coordinate, $R$ is the characteristic radius of the lens, and $n_1$ is the refractive index at $r=R$.
While Maxwell originally formulated the MFEL in three dimensions, its ray properties extend naturally to arbitrary spatial dimensions \cite{Luneburg_1944,Sahebdivan_2016}. In this work, we focus on a two-dimensional MFEL, illustrated in figure \ref{fig:MFEL_rays} \textbf{a}. 

We will further consider a particular version of the MFEL that was suggested by Shafer \cite{Shafer_1991,Shafer_1995} and Leonhardt \cite{Leonhardt_2009} who added a perfectly reflecting mirror at $r = R$. The combined effect of the mirror and the refractive index profile goes well beyond the focus-to-focus property of an elliptical mirror: here all rays emitted from any point source inside the lens are perfectly focused at their image point inside the lens.

The properties of the MFEL have attracted considerable attention, particularly around the possibility of perfect imaging not only in the ray description but also in wave optics (without the necessity of negative refraction) \cite{Leonhardt_2009,Makowski_2009,Tai_1958,Jagger_1992,Leonhardt_2010,Leonhardt_2010B,Leonhardt_2011B,Merlin_2011,Leonhardt_2015,Leonhardt_2015B,Alonso_2015,Abbasi_2019,Hu_2022,Chen_2022B,Astratov_2023}. Although the difficulty of engineering gradient index lenses has so far prevented an experimental demonstration in the visible part of the electromagnetic spectrum, several MFEL implementations have been experimentally realized in the GHz \cite{Abbasi_2019,Ma_2010,Ma_2011,Ma_2012,Dhouibi_2013}, THz \cite{Gabrielli_2012,Liu_2013}, and infrared regimes \cite{Bitton_2018}, while MFELs in other media have also been created, for example using elastic waves in metal plates of varying thickness \cite{Lefebvre_2015,Lefebvre_2023}. Beyond perfect imaging, there is interest in creating MFELs for applications ranging from the creation of integrated optical circuits, to the implementation of high fidelity coupling between separated quantum emitters \cite{Diekmann_2024,Perczel_2018}.

In this paper we report on the experimental realization of an analogue two-dimensional MFEL in a Bose–Einstein condensate (BEC) made from ultracold potassium-39 atoms.
In the acoustic (long wavelength) regime, phonons in a BEC obey a linear dispersion relation and are analogous to photons \cite{Pitaevskii_2016}. A refractive index profile like equation \eqref{eq:n_MFEL_pre} can be realized for these phononic excitations by a specific density profile (figure \ref{fig:MFEL_rays} \textbf{b}), which makes the phonon's speed of sound $c_s(r)$ spatially dependent.

Unlike most implementations for electromagnetic waves, where source and drain antennas are required to couple light in and out for observation, the acoustic realization in a BEC enables direct observation of the full phonon field in real time without the need for a source or drain.
In addition, BECs offer a high degree of \textit{in situ} tunability, and phononic wavepackets can in principle be created with arbitrary shape and spatial localization within the condensate and their propagation tracked over time. 
Furthermore, a BEC platform gives direct access to the quantum regime of the MFEL, since the phonons are genuine quantum excitations.

\section{Relation Between Maxwell Fish-Eye Optics and Condensate Acoustics}

\textit{\textbf{Optical metric and motion on a virtual sphere}} --- Waves traveling through a medium with a spatially varying refractive index $n(\textbf{r})$, such as light (photons) in a optical gradient-index lens or acoustic waves (phonons) in a BEC, have a spatially varying velocity $c(\mathbf{r})=c/n(\mathbf{r})$. Within the framework of geometric optics, ray trajectories correspond to null geodesics of an effective optical metric.
In a 2D plane with polar coordinates ($r,\varphi$) the ray geodesics obey the metric equation 
\begin{equation}
    0 = -dt^2 + \frac{1}{c^2(r)} (dr^2 + r^2 d\varphi^2).
    \label{eq:acoustic_metric_1}
\end{equation}
Substituting in the refractive index for the MFEL equation \eqref{eq:n_MFEL_pre} yields 
\begin{equation}
    0 = -dt^2 + \frac{1}{c^2} \left(\frac{2n_{1}}{1+\left(r/R\right)^2}\right)^2 \left( dr^2 + r^2 d\varphi^2 \right) \ .
    \label{eq:MFEL_metric_match}
\end{equation}

Under the stereographic projection $r = R\cot(\theta/2)$, this maps precisely to a metric describing geodesics (great circles) on the surface of a sphere with radius $R$ and a constant refractive index $n_{1}$,
\begin{equation}
 0 = -dt^2 + \frac{n_{1}^2}{c^2}  R^2 \left( d \theta^{2} + \sin^2 \theta \, d \phi^2 \right).
    \label{eq:metric_sphere}
\end{equation}
Here $(\theta,\phi)$ denotes the co-latitude and longitude angles on the sphere; the longitude on the sphere is equal to the azimuthal angle on the plane: $\phi=\varphi$. The stereographic projection establishes 
a one-to-one mapping between all points on the sphere (except the north pole) and a plane passing through its equator, as illustrated in figure \ref{fig:MFEL_rays}\,\textbf{c}.

This geometrical correspondence immediately explains the perfect imaging properties of the MFEL: rays or waves propagating in a physical lens with a spatially dependent refractive index $n(r)$ can be understood in terms of free propagation on the surface of a virtual sphere with a constant refractive index $n_{1}$. Rays emitted at any point on the sphere will follow great circles passing through that point and are perfectly focused at the antipodal point, corresponding to the 2D lens image point. The mirror positioned at $r=R$ corresponds to the equator of the sphere, $\theta=\pi/2$, and restricts the rays to the lower hemisphere.

Although the mapping above has been explicitly done in two dimensions, the argument can be generalized to show that a $d$-dimensional MFEL is conformal to the surface of a virtual sphere embedded in a $d+1$-dimensional Euclidean space \cite{Luneburg_1944,Leonhardt_2009}. The virtual sphere analogy has been previously utilized to simulate spherically curved spacetimes and the behavior of scalar fields therein by observing phonon propagation in a 2D BEC with the corresponding background density distribution \cite{Barcelo_2011,Tolosa_2022,Viermann_2022}.\\

\noindent
\textit{\textbf{Maxwell fish-eye lens in a Bose-Einstein condensate}} --- The condensate wavefunction $\Psi(\mathbf{r},t)$, describing the meanfield of a zero-temperature dilute BEC made from atoms of mass $m$ trapped in an external potential $V(\mathbf{r})$, obeys the Gross-Pitaevskii equation (GPE) 
\begin{equation}
    i\hbar\partial_t \Psi(\mathbf{r},t) = \left(-\frac{\hbar^{2}}{2m}\nabla^{2}+V(\mathbf{r}) +g \vert \Psi(\mathbf{r},t)\vert^{2}\right) \Psi(\mathbf{r},t)
    \label{eq:GPE}\ .
\end{equation}
The atomic density is given by  $\rho=\vert\Psi(\mathbf{r},t)\vert^2$ and $g$ characterizes the strength of interatomic contact interactions. In 3D the relation $g_{3\mathrm{D}}=4 \pi \hbar^2 a/m$ holds, where $a$ is the s-wave scattering length. To make the connection with the MFEL we shall consider the situation where the external trapping potential is cylindrically symmetric so that it is appropriate to describe the system using coordinates $(r,\varphi,z)$. Furthermore, we assume a harmonic trapping potential $V_{\perp}(z) = m\omega_z^2z^2/2$ along the $z$ direction tight enough to restrict the dynamics to the $x,y$-plane, giving rise to an effectively 2D condensate. The interaction constant of the quasi-2D BEC is $g=g_{\mathrm{3D}}/\sqrt{2\pi}l_{z}$, where $l_{z}=\sqrt{\hbar/m\omega_z}$ is the characteristic length scale of the harmonic oscillator, and the quasi-2D regime is satisfied when $g\rho\ll\hbar\omega_z$ \cite{Pitaevskii_2016}. 

In the acoustic regime~\cite{Pitaevskii_2016}, dispersion is absent and it has been shown that low-energy excitations (phonons) in the condensate obey the metric equation~\eqref{eq:acoustic_metric_1}, with the speed of light $c(r)$ replaced by local condensate speed of sound $c_{s}(r)=\sqrt{g\rho(r)/m}$ \cite{Barcelo_2011,Tolosa_2022}. We can establish the correspondence between the optical MFEL and the trajectories of phonons in a BEC by equating equations~\eqref{eq:acoustic_metric_1} to \eqref{eq:MFEL_metric_match}, and identifying $c/2n_1=c_0 = c_s(0)$. From this we obtain the required density profile such that phonons in the condensate experience the MFEL refractive index:
\begin{equation}
    \rho(r)=
    \rho_{\mathrm{0}}\left(1+\frac{2r^{2}}{R^{2}}+\frac{r^{4}}{R^{4}}\right).
    \label{eq:rho(r)_MFEL}
\end{equation}
Here $\rho_{\mathrm{0}} =\rho(0)=m c_{0}^{2}/g$ is the density at the center of the condensate, while the density at the edge is given by $\rho(R)=4\rho_{\mathrm{0}}$. An experimental realization of this density distribution is depicted in figure ~\ref{fig:MFEL_rays}\,\textbf{b}. In the Thomas-Fermi limit, the corresponding potential for such a stationary density distribution obeys
\begin{equation}
    \mu = g\rho(r) + V(r),
    \label{eq:Euler_stationary}
\end{equation}
where the chemical potential $\mu$ is constant. If we impose the condition $V(0)=0$, equation \eqref{eq:Euler_stationary} implies $\mu=g\rho_{\mathrm{0}}$. Substituting this and equation \eqref{eq:rho(r)_MFEL} into equation \eqref{eq:Euler_stationary} and rearranging for $V(r)$ yields 

\begin{equation}
 V(r) =-\frac{1}{2} m \omega^2 \left(r^{2}+\frac{r^{4}}{2R^{2}}\right),
    \label{eq:V_MFEL}
\end{equation}
where we have introduced $\omega$ via $m\omega^{2}/2=2g\rho_{\mathrm{0}}/R^{2}$.
An infinite cylindrical potential barrier at $r = R$ plays the role of the mirror for the MFEL.
In the full GPE theory the condensate density approaches zero as $r\rightarrow R$ over the characteristic healing length scale $\xi(r) = \hbar / \sqrt{2 m g\rho(r)}$. For phonons with wavelength greater than the healing length this effectively realizes a Neumann boundary condition, a prerequisit for perfect reflection \cite{Ratzel_2018}.
Having established the required atomic density profile of the BEC, the condensate can now be used to simulate an optical MFEL. We can derive the total time $T$ it takes a phonon to propagate from a given location in the MFEL to its associated antipodal (image) point by integrating the metric equation \eqref{eq:acoustic_metric_1} with local speed of sound $c_{s}(r)$. This yields
\begin{equation}
    T = 2\int_{0}^R\frac{dr}{c_s(r)} =\frac{R}{2c_{0}}\int_{-\pi/2}^{\pi/2}d\theta = \frac{\pi R}{2 c_{\mathrm{0}}} =\frac{\pi}{\omega}.
    \label{eq:total_time}
\end{equation}
Although the first integral can be solved directly using equation \eqref{eq:rho(r)_MFEL}, the stereographic transformation $r = R\cot(\theta/2)$ gives a clear physical interpretation in terms of the virtual sphere via the second integral: regardless of the source location in the MFEL, any given ray travels half the circumference on the surface of the sphere with constant speed $ c_{\mathrm{0}}$, as depicted in figure \ref{fig:MFEL_rays} \,\textbf{c}. The factor of $1/2$ comes from the conformal factor of the stereographic projection map. 

The question of optimal focusing in a Maxwell fish-eye configuration is closely connected to the discussion of sub-wavelength resolution and perfect imaging in a MFEL in the context of electromagnetic waves \cite{Leonhardt_2009}. 
In that case, the concept of perfect imaging without a wavelength limit is a theoretical construct that emerges as an interpretation of a sharp sub-wavelength feature of the Greens function, which is a stationary monochromatic solution of the Helmholtz equation for a point-like source and also a point-like coherent drain localized at the image point.
This makes the phenomenon very elusive — for example non-monochromatic sources and the absence of a coherent drain already precludes perfect imaging as defined in \cite{Leonhardt_2009}.
Furthermore, these concepts can not be directly applied to the phononic excitations of a BEC, where the metric description is only valid in the long-wavelength limit.
In our system, the healing length $\xi\approx0.64\,\mu$m sets the scale at which the dispersion deviates significantly from the linear regime.

\section{Numerical and Experimental Verification}

\textit{\textbf{Experimental platform}} --- Our experimental system is based on a quasi-2D BEC of $6\times10^4$ atoms of potassium-39. 
The BEC is tightly confined in the vertical direction with a harmonic, repulsive dipole trap with a trap frequency of $\omega_z=2\pi\times 1.5\,$kHz, limiting the BEC's dynamics to the horizontal plane.
The in-plane confinement is created with a second repulsive dipole potential whose shape can be controlled with a Digital-Micromirror Device (DMD), making it possible to create various trap shapes such as box traps and very general continuous trapping geometries, like that required for the MFEL. The potential is optimized using an iterative feedback algorithm based on the method of reference \cite{Guillaume_2020} to produce the target atomic density of the MFEL, equation \eqref{eq:rho(r)_MFEL}.
The inter-atomic interactions can be tuned using a magnetic Feshbach resonance at 561$\,$G and are fixed to $200\,$a$_0$ for the results discussed here. 
The speed of sound at the center of the MFEL $c_0\approx1.8 \pm 0.1\,\mathrm{\mu m/ms}$ is measured by preparing a wave packet in the center of the condensate and tracking the wave front position over time.
Two-dimensional, \textit{in situ} density distributions are read out using high resolution absorption imaging \cite{Maurus2021}. Setting $R=36\pm 0.5$\,$\mu$m, the density profile of our effective MFEL is displayed in figure \ref{fig:MFEL_rays}\,\textbf{b}. All experimental data shown are averaged over approximately 30 realizations and groups of \(2\times2\) camera pixels were binned, where one camera pixel has a size of $(0.455\,\mu$m$)^2$ in the atom plane, corresponding to an optical resolution $\sigma\sim0.4\,\mu$m of the imaging objective.\\

\begin{figure}[!t]
    \centering
    \includegraphics[width=1\linewidth]{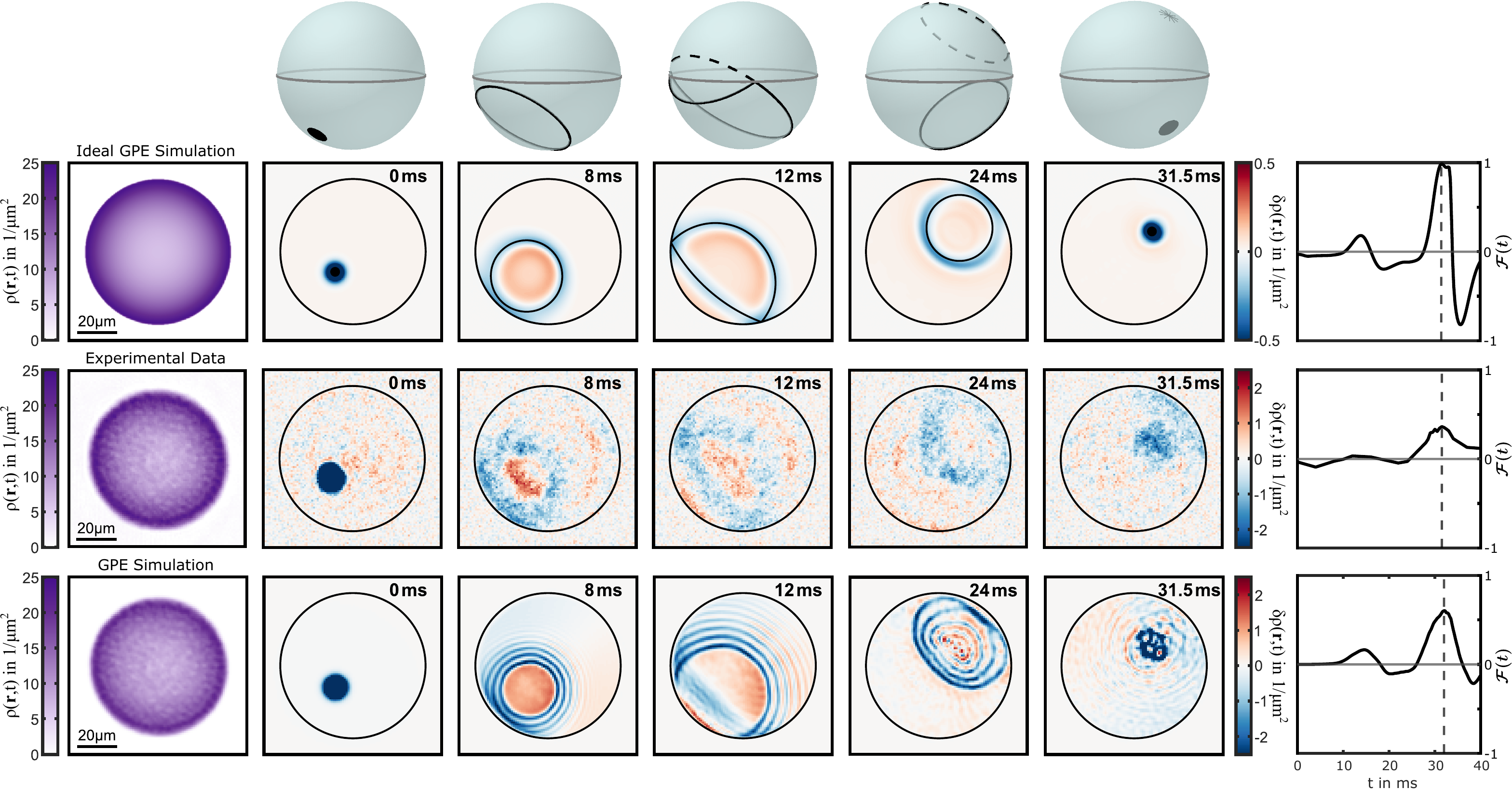}
    \caption{Snapshots in time of our effective Maxwell fish-eye lens focusing dynamics: analytic wave-fronts represented on the surface of a sphere (first row), idealized Gross-Pitaevskii equation simulation (second row), experimental results (third row), and again simulation but now using the initial $t=0$ experimental background density (fourth row). The left column shows the initial density distribution of the lens. The middle columns show the density difference for different times, $\delta\rho(\mathbf{r},t)=\rho(\mathbf{r},t)-\rho_{BG}(\mathbf{r})$, where $\rho_{BG}(\mathbf{r})$ is the initial density without a perturbation. For the idealized simulation a perturbation depth of $15\%$ was used, while in the other cases a depth of $70\%$ was used. The right column shows the fidelity (equation \ref{eq:fidelity}) between the density difference at a given time $\delta \rho(\mathbf{r},t)$ versus the expected density difference if perfect focusing were to occur at the antipodal point, $\delta\rho(-\mathbf{r},0)$. The standard deviation of the mean, obtained by evaluating the numerator of the fidelity for individual realizations, is smaller than the plotted line width. The maximum overlap is at $t = 31.5\,\mathrm{ms}$ (vertical dashed line), in agreement with the predicted time of $T = 31.4 \pm 1.8\,\mathrm{ms}$ obtained via equation \eqref{eq:total_time}.}
    \label{fig:dynamics1}
\end{figure}

\noindent
\textit{\textbf{Focusing dynamics}} --- To demonstrate focusing in our lens we create an approximately $11$\,$\mu$m in diameter localized excitation in the condensate by initially adding a potential that creates a density indent at $\mathbf{r}_{0}=(x_{0},y_{0})=(-10.5,-9.6)$\,$\mu$m. By quickly removing the potential that creates the indent the excitation wavepacket is allowed to propagate freely through the condensate, reflect from the boundary mirror, and focus at approximately time $T$ at the antipodal point, $-\mathbf{r}_{0}$. Ideally the indent should be sufficiently small in amplitude and gently sloped at its edges in order to only excite purely phononic waves with linear dispersion, $\Omega=c_{s}(r)k$. 
Non-phononic (large $k$) waves obey the Bogoliubov dispersion relation  \cite{Pitaevskii_2016} and have a wavelength dependent speed of sound. Thus, they do not perfectly follow trajectories governed by the MFEL ray geodesics, which smears the idealized focusing point. Experimentally, a similar effect was observed in references \cite{Lefebvre_2015,Lefebvre_2023} for elastic waves in a metal plate MFEL, where the dispersion relation also becomes nonlinear at higher $k$.
In our experiment the density indent is created using a repulsive dipole potential. It is important to note that there is a tradeoff between the depth of the initial density indent and signal to noise ratio; finite temperature, finite healing length, and DMD resolution effects can all result in deviations from perfect focusing. 

Figure \ref{fig:dynamics1} depicts the experimental and numerical focusing dynamics via snapshots in time of the density difference $\delta\rho(\mathbf{r},t)=\rho(\mathbf{r},t)-\rho_{BG}(\mathbf{r})$, where $\rho_{BG}(\mathbf{r})$ is the initial $t=0$ density without a perturbation. The propagation of the wave front can be understood as a light cone on the surface of the sphere, which appears bent when projected onto the plane using the stereographic projection (see the first and second rows of figure \ref{fig:dynamics1}).

The numerical simulations shown in the second row of figure \ref{fig:dynamics1} are obtained by solving the GPE \eqref{eq:GPE} using the split-step-Fourier method. We prepare an initial state by first propagating the GPE in imaginary time in the presence of potential (\ref{eq:V_MFEL}), in order to find the ground state of the system. We also verify that the experimental parameters give $\mu/\hbar\omega_z\approx 0.18$, and $\mu/\hbar\omega\approx 5.2$ indicating that we are reasonably-well described by the quasi-2D and in-plane Thomas-Fermi descriptions, respectively~\cite{Pitaevskii_2016}. At $t=0$ we then impose a small density indent that is gaussian in form, selecting parameters to ensure gently sloped edges: standard deviations $\sigma_{x}=\sigma_{y}\approx3.5\mu\mathrm{m}$ and amplitude $A\approx0.195\mu\mathrm{m}^{-1}$. This results in a density dip that is about $15\%$ of the local background density. We characterize the focusing quality of the MFEL via the perturbation fidelity 
\begin{equation}
\mathcal{F}(t) = \langle \delta \rho(-\mathbf{r}, 0)\: \delta \rho(\mathbf r, t) \rangle
/\sqrt{\langle \delta \rho(-\mathbf{r}, 0)^2\rangle\langle \delta \rho(\mathbf r, t)^2 \rangle}, \label{eq:fidelity}
\end{equation}
which corresponds to the normalized overlap between the density difference at any time $t$ and an inverted image of the initial perturbation $\delta \rho(-\mathbf{r}, 0)$, where $\langle...\rangle=\int...\,\mathrm{d}\mathbf{r}$. Perfect focusing by a lens at time $T$ then implies $\delta \rho(\mathbf{r}, T)=\delta \rho(-\mathbf{r}, 0)$, and hence we expect $\mathcal{F}(T)=1$ in the ideal case.
The GPE goes beyond the Thomas-Fermi description and fully includes dispersion, accounting for finite size effects even down to the scale of the healing length. These effects include the finer features of the initial indent and also the boundary at the mirror where the density goes to zero on the scale of the healing length. In the second row of figure \ref{fig:dynamics1} we refer to this case as the idealized GPE simulation since, although it is a full solution of the GPE, it does not include experimental imperfections.

In the experiment itself, a density dip of about 70\% of the local background density is necessary because phonon damping in our system is significant, leading to a low signal-to-noise ratio at later times. Although such a large perturbation is not in the linear regime of the phonon description, the amplitude decreases quickly as the wavefront coherently spreads outwards, rapidly giving rise to a linear perturbation. Additionally, high-$k$ excitations, which are not expected to follow the MFEL geodesics, are damped out well before the focusing time. Also, the finite resolution of the DMD limits how sharp the boundary mirror potential at $r=R$ can be; the width at the edge is about $3.4\,\mu\mathrm{m}$, corresponding to the distance over which the density decreases from 90\% to 10\% of the maximum background density. This is consistent with the expected healing length at the edge ($\xi(R)\approx 0.35\,\mu\mathrm{m}$), convolved with a Gaussian of standard deviation $\approx 0.7\,\mu\mathrm{m}$ which combines the effects of the DMD resolution ($\sigma\approx0.2\,\mu\mathrm{m}$), the optical imaging resolution ($\sigma\approx0.4\,\mu\mathrm{m}$) and a blur effect ($\sigma \approx0.5\,\mu\mathrm{m}$) due to the atoms diffusing during the $10\,\mu\mathrm{s}$ imaging pulse.
Despite these deviations from the idealized regime, there is a clear experimental signal of focusing as shown in the third row of figure \ref{fig:dynamics1}, and the agreement with theory is also apparent at earlier times. For the experimental data, at the focusing time we achieve a maximum fidelity of approximately $\mathcal{F}(t) \approx  0.36$. 
To quantify the broadening and amplitude decrease of the wavepacket, we fitted a Gaussian profile at $t = 0\,\mathrm{ms}$ and $t = 31.5\,\mathrm{ms}$ to $x$- and $y$-cuts of the density difference. We observe that the wavepacket broadens by a factor of $1.65$ and the total integral decreases by a factor of $\sim 2$. This broadening and decrease in amplitude can be explained by the finite lifetime of phonons, where sharper features are damped more quickly. Numerical simulations showed that a smoother edge of the condensate does not lead to significant broadening of the wavepacket.

\begin{figure}[!t]
    \centering
    \includegraphics[width=1\linewidth]{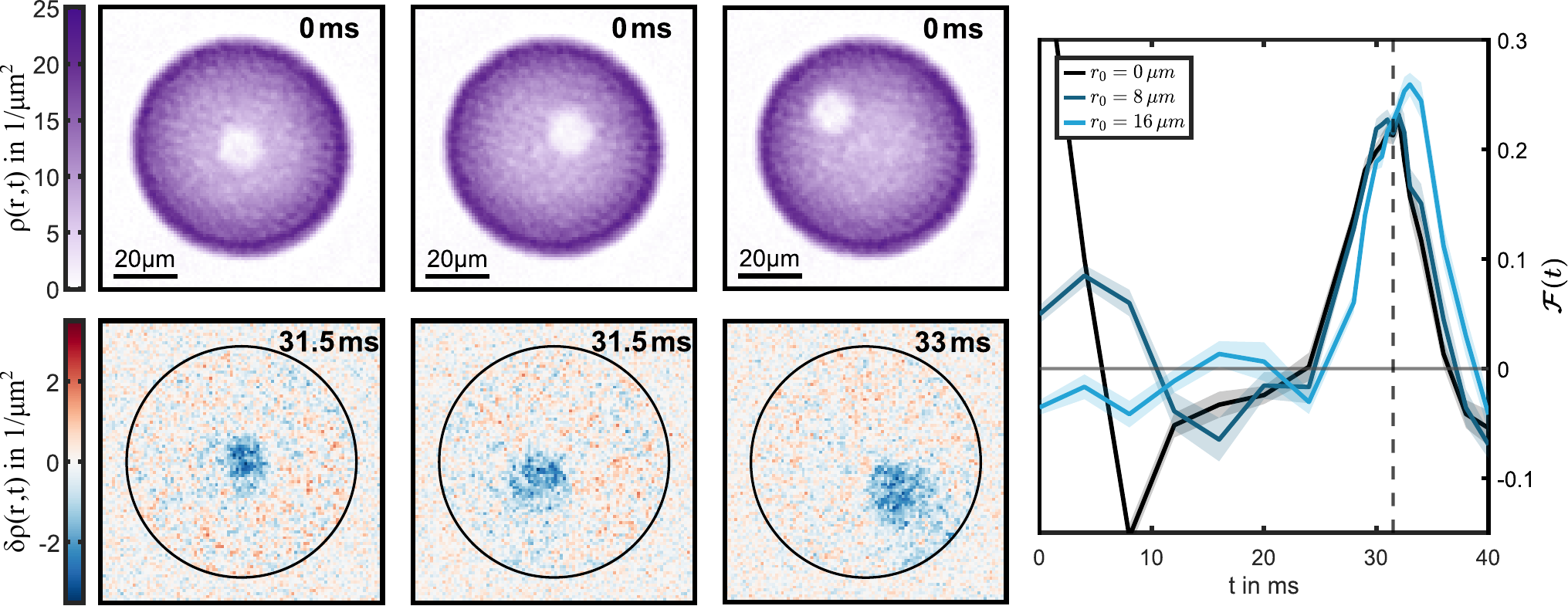}
    \caption{Focusing fidelity for different initial wavepacket positions $r_{0}$: the density at $t = 0\,\mathrm{ms}$ including the wavepacket indent (first row), the density difference at the time of maximum fidelity (second row), and fidelity (eq. \ref{eq:fidelity}) over time for all initial positions (rightmost image). The shading indicates the standard deviation of the mean, obtained by evaluating the numerator of the fidelity for individual realizations. The fidelity shows a clear peak around $t = 31.5\,\mathrm{ms}$ (vertical dashed line). For larger initial displacements from the central position the peak shifts slightly towards later times. }
    \label{fig:dynamics3}
\end{figure}

To determine which deviations in the experimental focusing come from the idealized GPE theory and which arise from a non-optimal MFEL density, wavepacket size, or other experimental sources, we also performed a GPE simulation that used the experimentally measured density without a perturbation at $t=0$ in order to start from the same initial density profile as in the experiment. The results are shown in the fourth row of figure \ref{fig:dynamics1}. The chemical potential $\mu=\rho_0 g$ in the GPE simulation was matched to the experimental value by matching the speed of sound at the center, $c_0 \approx 1.8\,\mu\mathrm{m/ms}$. To achieve this, the interaction strength in the numerical simulation was increased by approximately 30\%. Deviations in the speed of sound predicted by the local density approximation versus the experimentally measured speed of sound arise from a systematic discrepancy in the atom number estimation obtained from the absorption imaging.
From the density difference obtained in the simulation, it becomes clear that phonons outside the linear regime of the dispersion are excited. In this simulation a fidelity of approximately $\mathcal{F}(t)\approx0.58$ is observed, indicating that deviations from the optimal potential, as well as the wavepacket size and amplitude, account for part of the discrepancy with the idealized simulation. The additional reduction in fidelity observed in the experiment can be attributed to the decrease in signal amplitude arising from finite temperature effects and a limited phonon lifetime, or other experimental imperfections that are not included in the GPE model.

To demonstrate that antipodal focusing in the MFEL occurs at the predicted time regardless of the initial source location $r_{0}$, we performed wavepacket propagation from three different initial positions, varying both radial and angular coordinates. The initial density (first row) at $t = 0$ as well as the density difference (second row) at the time $t = 31.5\,\mathrm{ms}$ are shown in figure \ref{fig:dynamics3}. Here, only 20 realizations were averaged. For all three cases, a clear focusing signal is observed, with a fidelity ranging from 0.22 to 0.25 at around $t = 31.5\,\mathrm{ms}$ as shown in the right graph of figure \ref{fig:dynamics3}. The apparent shift towards larger times for larger initial displacements can be numerically verified to stem from the nonlinearity of the GPE, where a deep density perturbation leads to a disturbance in the metric. This effect is stronger in regions of higher density, e.g. towards the edge, where the wavepacket displaces more atoms.

\section{Concluding Remarks}

By establishing the correspondence between light rays in an optical MFEL and the trajectories of phonons in a quasi-2D BEC, we determined the required density profile and corresponding trapping potential such that phonons traveling through the BEC simulate photons traveling through a MFEL with a mirror placed at its edge. We experimentally and numerically verified this by studying the point-to-point focusing dynamics of an initial perturbation in the condensate. Despite real-world imperfections (thermal noise, finite healing length, DMD resolution, etc.) that reduce the focusing fidelity relative to the idealized theory, we still observe a clear focusing signal in the experiment at the analytically predicted focusing time.

Our work not only adds a new experimental realization of a MFEL, but also makes the wave dynamics in a MFEL observable in real time. It would also be interesting to explore the MFEL mode structure and the possibility of creating other types of effective lenses; the procedure we follow can in principle be followed to simulate other perfect focusing instruments with different applications, e.g., the Lüneberg lens, Eaton lens, or generalized MFELs \cite{Tyc_2011,Eskandari_2019,Xu_2019,Yin_2020}.
An alternative question is whether it is possible to engineer effective curved geometries for vortices or vortex dipoles, rather than for phonons.
The connection between the motion of vortex dipoles, refractive indices, and curved geometries has been studied before in related contexts \cite{Cawte_2019,Bereta_2021,Caracanhas_2022}. Finally, in analogy to the high fidelity coupling of separated quantum emitters in a MFEL \cite{Diekmann_2024,Perczel_2018}, we anticipate that impurity atoms in the BEC located at antipodal points will experience strong interactions in the form of long-range phonon-mediated potentials. Whereas impurity atoms in 3D uniform BECs (and superfluids more generally) exchange virtual excitations that lead to an exponentially decaying Yukawa-type interaction \cite{Bijlsma00,Naidon2018,Drescher23}, by contrast we expect oscillating impurities in a MFEL to have essentially infinite range interactions with partner impurities located at antipodal points.

\section*{Acknowledgments}
We thank Marcel Kern and Nikolas Liebster for supporting the experimental activities at the beginning of the project and Hung Nguyen, Sriram Sundaram, and Mark Dennis for useful conversations.
This work is supported by the Deutsche Forschungsgemeinschaft (DFG, German Research Foundation) under Germany’s Excellence Strategy EXC 2181/1 - 390900948 (the Heidelberg STRUCTURES Excellence Cluster) and under SFB 1225 ISOQUANT - 273811115, as well as the Natural Sciences and Engineering Research Council of Canada (NSERC) [Reference No. RGPIN-2025-06703].

\section*{Data Availability Statement}
The data cannot be made publicly available upon publication because they are not available in a format that is sufficiently accessible or reusable by other researchers. The data that support the findings of this study are available upon reasonable request from the authors.

\section*{Competing Interests Statement}
The authors declare no competing interests.

\appendix

\section{Ray Hamiltonian and circular rays}
\label{app:ray_hamiltonian}

In this appendix we summarize the geometrical optics (eikonal) description of rays in a medium with a spatially varying wave speed, and then specialize to the case of the MFEL to show that in this case the ray trajectories are circles (or straight lines in the infinite-radius limit).
We start from the local dispersion relation $\Omega = c (\bm r) |\bm k|$ in a static medium and use it as a Hamiltonian to obtain the ray equations of motion. These can be combined to express the dynamics in terms of a complex ordinary differential equation, which is precisely the geodesic equation for the spherical metric written in stereographic coordinates. Since spherical geodesics are great circles, their images under a conformal stereographic mapping  (i.e., planar ray trajectories) are circles (or straight lines) in the plane.

In a static 2D background, the local frequency $\Omega$ is conserved along the rays. A natural choice of ray Hamiltonian is therefore the dispersion relation
\begin{equation}
    \mathcal{H}(\bm r,\bm k)\equiv \Omega(\bm r,\bm k) = c(\bm r)\,\abs{\bm k},
    \label{eq:app_rayH}
\end{equation}
which is the standard Hamiltonian form of geometrical optics \cite{Leonhardt_2010,Born_1980}.
Using physical time $t$ as the evolution parameter, Hamilton's equations read
\begin{align}
    \dot{\bm r} &= \pdv{\mathcal{H}}{\bm k} = c(\bm r)\,\frac{\bm k}{\abs{\bm k}} \equiv c(\bm r)\,\hat{\bm k},
    \label{eq:app_ham_r}\\[0.3em]
    \dot{\bm k} &= -\pdv{\mathcal{H}}{\bm r} = -\abs{\bm k}\,\nabla c(\bm r).
    \label{eq:app_ham_k}
\end{align}
Eliminating $\bm k$ gives a closed second-order equation for the ray path:
\begin{equation}
    \ddot{\bm r}
    = -c (\bm r) \,\nabla c (\bm r)
      + 2\,\frac{\dot{\bm r}}{c (\bm r)}\,(\dot{\bm r}\cdot\nabla c (\bm r)),
    \label{eq:app_rddot_full}
\end{equation}
together with the constraint $\abs{\dot{\bm r}} = c(\bm r)$. The MFEL refractive index of equation \eqref{eq:n_MFEL_pre} with $r = \abs{\bm{r}}$ sets the wave-speed via 
\begin{equation}
    c(r) = \frac{c}{n(r)} = \frac{c}{2n_{1}}\,\left(1 + \frac{r^{2}}{R^{2}}\right),
    \label{eq:app_c_fisheye}
\end{equation}
where $c$ is the vacuum wave-speed.
It is interesting to note that the ray Hamiltonian in the 2D MFEL case has the following form:
\begin{equation}
    \mathcal{H}_{\textrm{MFEL}}(\bm r,\bm k) = \frac{c}{2n_{1}}\,\left(1 + \frac{r^{2}}{R^{2}}\right) \, \abs{\bm k}.
\end{equation}
Substituting \eqref{eq:app_c_fisheye} into \eqref{eq:app_rddot_full} gives
\begin{equation}
    \ddot{\bm r} = - \frac{c^{2}}{2n_{1}^{2}R^2}\,(1 + \frac{r^2}{R^2})\,\bm r + \frac{4(\bm r \cdot \dot{\bm r})}{R^2 + r^2}\,\dot{\bm r}, \qquad \abs{\dot{\bm r}}=\frac{c}{2 n_{1}}(1+\frac{r^2}{R^2}).
    \label{eq:app_rddot_fisheye}
\end{equation}
Now we introduce the following dimensionless variables
\begin{equation}
    \bm u \equiv \frac{\bm r}{R}, \qquad \tau \equiv \frac{c}{2n_1 R} t,
\end{equation}
so $u = \abs{\bm u}$ and primes denote $d/d\tau$. 
Then \eqref{eq:app_rddot_fisheye} becomes
\begin{equation}
    \bm u'' = -2\,(1+u^2) \bm u + \frac{4(\bm u \cdot \bm u')}{1+u^2} \bm u', \qquad \abs{\bm u'} = 1 + u^2.
    \label{eq:app_u_primes}
\end{equation}
Defining the complex coordinate $z = u_x + i u_y$ (so $u^{2}=|z|^{2}$) and using $\bm u \cdot \bm u'=\Re(\bar z\,z')$, equation \eqref{eq:app_u_primes} becomes
\begin{equation}
    z'' = \frac{4\,\Re(\bar z \, z')}{1 + \abs{z}^{2}} \, z' - \frac{2\,\abs{z'}^{2}}{1+\abs{z}^{2}}\,z,
\end{equation}
along with the constraint $|z'|=1+|z|^2$.
The numerator simplifies as $4 \Re(\bar z\, z')\, z' -2\,\abs{z'}^{2}\,z = 2\,\bar z\,(z')^{2}$, hence
\begin{equation}
    z'' = \frac{2\,\bar z}{1 + \abs{z}^{2}}\,(z')^{2}.
    \label{eq:app_fisheye_ray_equals_geodesic}
\end{equation}

Equation \eqref{eq:app_fisheye_ray_equals_geodesic} is the stereographic-form geodesic equation on $\mathbb{S}^{2}$, so the rays are images of great circles and hence trace circles (or straight lines) in the plane. 
Moreover, any great circle is obtained from the equator by a rotation of the sphere, and stereographic projection sends such rotations to fractional linear (M\"obius) maps of the form
\begin{equation}
    z \mapsto \frac{\alpha z+\beta}{-\bar\beta z+\bar\alpha},
    \qquad |\alpha|^2+|\beta|^2=1,
    \label{eq:su2_mobius_map}
\end{equation}
i.e.\ the standard $SU(2)$ action on the Riemann sphere (see, e.g., \cite{Needham_2023}).
Applying \eqref{eq:su2_mobius_map} to the equator solution $z_{\rm eq}(\tau)=e^{2i \tau}$ gives the general trajectory
\begin{equation}
    z(\tau)= \frac{\alpha \, e^{2i \tau} + \beta}{-\bar{\beta} \, e^{2i \tau} + \bar{\alpha}},
    \label{eq:mobius_general_solution2}
\end{equation}
with $\abs{\alpha}^2 + \abs{\beta}^2 = 1$.
To see explicitly that \eqref{eq:mobius_general_solution2} traces a circle (or line), we set $w = e^{2i\tau}$ and rewrite
\begin{equation}
    z = \frac{\alpha w + \beta}{-\bar\beta w + \bar\alpha}
    \quad \Longleftrightarrow \quad
    w = \frac{\bar\alpha z - \beta}{\alpha + \bar\beta z}.
\end{equation}
So $|w| = 1$ gives the $\tau$-independent locus
\begin{equation}
    |\bar\alpha z - \beta|^{2} = |\alpha + \bar\beta z|^{2}
    \;\Longrightarrow\;
    (|\alpha|^{2} - |\beta|^{2})(|z|^{2} - 1) - \alpha\beta\,z - \bar\alpha\bar\beta \, \bar z = 0.
\end{equation}
Using $z \equiv u_x + i u_y$ from the definition above, the locus becomes
\begin{equation}
    (|\alpha|^{2} - |\beta|^{2})(u_x^{2} + u_y^{2} - 1) - 2 \, \Re(\alpha\beta) \, u_x + 2\,\Im(\alpha\beta) \, u_y = 0,
\end{equation}
which is the standard quadratic form of a circle when $|\alpha|\neq|\beta|$, and degenerates to a straight line when $|\alpha|=|\beta|$.

\printbibliography

\end{document}